\documentclass[a4paper]{article}
\usepackage{amsmath}
\usepackage{amssymb}
\usepackage{mathtext}
\usepackage[T2A]{fontenc}
\usepackage{latexsym}
\usepackage[utf8]{inputenc}
\usepackage[english,russian]{babel}

\ifx\pdfoutput\undefined
\usepackage{graphicx}
\else
\usepackage[pdftex]{graphicx}
\fi

\makeatletter
\usepackage{amsbib}
\renewcommand{\@biblabel}[1]{#1.}
\makeatother

\usepackage{cite}
\usepackage{authblk}
\usepackage{multirow}
\newcommand{\Keywords}[1]{\par\noindent{\small{\em \textbf{Keywords}\/}: #1}}

\title{Continuum Percolation on Disoriented Surfaces: the Problem of Permeable Disks on a Klein Bottle}
\author{V. D. Borman, A. M. Grekhov, V. N. Tronin, I. V. Tronin \thanks{IVTronin@mephi.ru}}
\affil{National Research Nuclear University MEPhI}
\date{Kashirskoe sh. 31, Moscow, 115409 Russia}

\begin{document}

\maketitle

\begin{abstract}
The percolation threshold and wrapping probability $R_{\infty}$ for the two-dimensional problem of continuum percolation on the surface of a Klein bottle have been calculated by the Monte Carlo method with the Newman--Ziff algorithm for completely permeable disks. It has been shown that the percolation threshold of disks on the Klein bottle coincides with the percolation threshold of disks on the surface of a torus, indicating that this threshold is topologically invariant. The scaling exponents determining corrections to the wrapping probability and critical concentration owing to the finite-size effects are also topologically invariant. At the same time, the quantities $R_{\infty}$ are different for percolation on the torus and Klein bottle and are apparently determined by the topology of the surface. Furthermore, the difference between the $R_{\infty}$ values for the torus and Klein bottle means that at least one of the percolation clusters is degenerate.
\end{abstract}

\section{Introduction. Formulation of the problem}

Problems of the percolation theory are widely used both in fundamental physics~\cite{perc:book_kesten, perc:grimmet, perc:2d_conformal, perc:cardy_arxiv, perc:cardy, perc:cft, perc:cardy_delfino, perc:delfino, perc:smirnov, perc:dimers, perc:delfino2, perc:lorenz_ziff, perc:moloney, perc:2dcont, perc:moloney2003, perc:ziff_PRE_2010, perc:wang2013, perc:cohen, perc:PRL_85, perc:PRE_78, perc:janson} and in applications. The fundamental interest in the percolation theory is associated with studies of the conformal field theory, probability theory, statistical physics, and theory of random graphs~\cite{perc:book_kesten, perc:grimmet, perc:cardy_arxiv, perc:cardy_delfino, perc:delfino, perc:dimers, perc:delfino2, perc:lorenz_ziff, perc:moloney, perc:cohen, perc:PRL_85, perc:PRE_78, perc:janson}. In applied problems, models of the percolation theory are used to describe granulated and composite materials~\cite{perc:odagaki, perc:tobochnik, perc:bondt}, propagation of illnesses~\cite{perc:grassberger, perc:newman_PRE_66}, and reliability of networks~\cite{perc:cohen, perc:PRL_85, perc:PRE_65, perc:PRL_108}. Models of the percolation theory are the main models for the description of processes in structurally disordered porous media filled with liquids and gases~\cite{perc:machta91, perc:superfluid, perc:sahimi_revmodphys}. The states and properties of such systems have been actively studied in recent years~\cite{perc:PRE_90, perc:PRE_91, perc:softmatter, perc:PRE_67, perc:PRL_87, perc:lefevre, perc:langmuir, perc:rigby, perc:xu, perc:han, perc:french, perc:grekhov13, perc:grekhov15, perc:borman_dispersion, perc:borman_JETP, perc:borman_PLA, perc:borman_PRE}. Phenomena beyond the traditional notions were detected for such systems. In particular, the disordered system of pores of a porous medium filled with a nonwetting liquid can undergo a dispersion transition, when the nonwetting liquid transits to an effectively <<wetting>> state at the variation of the temperature and degree of filling of the porous medium~\cite{perc:borman_dispersion, perc:borman_JETP}. The state of such a system can be nonergodic with an anomalously slow relaxation of nonequilibrium states~\cite{perc:borman_PLA, perc:borman_PRE}.

A physical reason for the dispersion transition is the appearance of the collective <<multiparticle>> interaction of liquid nanoclusters in neighboring pores with various sizes~\cite{perc:borman_PLA, perc:borman_PRE} owing to the formation of the ground state in the form of the percolation cluster of liquid-filled pores inside the percolation cluster of empty pores of the porous medium. The relaxation of the system is a process of its successive transition through the local maxima of the energy of local configurations of clusters of empty and filled pores and depends on the degeneracy of the ground state~\cite{perc:borman_PLA, perc:borman_PRE}, which is determined by the percolation cluster of filled pores. It is known that continuum percolation models~\cite{perc:machta91, perc:superfluid, perc:sahimi_revmodphys, perc:borman_PLA, perc:borman_PRE, perc:kheifets_neimark} are used to describe the behavior of a liquid under the confinement conditions in disordered porous media. For this reason, the problem of the universality and properties of the percolation cluster in the case of continuum percolation arises, including the problem of the degeneracy of the percolation cluster in this case. We note that similar problems of the universality of various characteristics in two-dimensional percolation theory were previously studied both analytically and numerically~\cite{perc:lorenz_ziff, perc:moloney, perc:2dcont, perc:moloney2003} for a sphere~\cite{perc:lorenz_ziff}, a torus~\cite{perc:2dcont}, and a M\"obius strip~\cite{perc:moloney}.

In this work, the problem of the possible degeneracy of the percolation cluster in the case of continuum percolation is solved by the numerical Monte Carlo simulation. To this end, we calculate the wrapping probability $R_{\infty}$ in the two-dimensional problem of percolation on a Klein bottle and compare with the results of the known calculation of percolation on the torus~\cite{perc:2dcont}. It is shown that the percolation thresholds and scaling exponents for the torus and Klein bottle coincide with each other, whereas the quantities $R_{\infty}$ are significantly different. The last circumstance can be due to the degeneracy of the percolation cluster on one or both surfaces.

\section{Method of the calculation}

The Monte Carlo calculations were performed on square systems of various sizes with the Newman--Ziff algorithm~\cite{newmanziff}. This algorithm is currently the fastest algorithm of the search for the percolation threshold and ensures the calculation time proportional to the number of objects in a system (e.g., the calculation time in the Hoshen--Kopelman algorithm~\cite{hoshenkopelman} is proportional to the \textit{square} of the number of objects in the system). The main idea of the algorithm is that objects are added to the system one-by-one; after that, the conditions of the overlapping of an added object with already existing objects are verified, objects are joined into clusters, and, finally, percolation conditions are verified. The linear dependence on the number of objects is reached owing to the use of the \textit{union find with path compression} algorithm~\cite{comp:algorithms, comp:knuth_algo} for the joining of clusters.

In this work, completely permeable disks with the unit diameter $d=1$ are used as objects. According to the Newman--Ziff algorithm, the following actions are performed in each calculation iteration:
\begin{enumerate}
 \item The addition of a disk to the system.
 \item The test of the conditions of overlapping of the added disk with the disks already existing in the system and joining of disks into clusters.
 \item The test of the percolation conditions.
\end{enumerate}
The addition of a disk to the system is reduced to the generation of two random numbers determining the position of its center. The random numbers were generated according to the Mersenne--Twister algorithm having a period of $2^{19937}-1$.

The test of the conditions of overlapping of the added disk with the disks already existing in the system is complicated because the number of neighbors of the disk in the case of continuum percolation is unknown. In order to ensure the effective operation of the algorithm, we used an approach proposed in~\cite{perc:2dcont}; the essence of this approach is as follows. The calculation domain (a square with a certain size) is divided into squares with a side equal to the diameter of a disk, i.e., with a unit side. A disk located in a certain cell can overlap only with disks located in the same cell or in eight cells surrounding the given cell. Thus, the condition of overlapping is verified only for disks in the given cell and eight neighboring cells.

Disks are joined into clusters according to the Newman--Ziff algorithm~\cite{newmanziff}. For the subsequent test of the percolation conditions, the shifts over two axes from the given disk to the root disk of the cluster are calculated in the process of joining. This path is formally a floating-point number, but it is easier to use the method proposed in~\cite{perc:2dcont} and to calculate the path from the square cell containing the disk to the cell of the root disk, rather than the path from the disk to the root of the cluster. This method allows using integers in the calculation of the path and in the test of the percolation conditions.

\begin{figure}[h!]
\begin{minipage}[h]{0.49\linewidth}
\center{\includegraphics[width=1.0\linewidth]{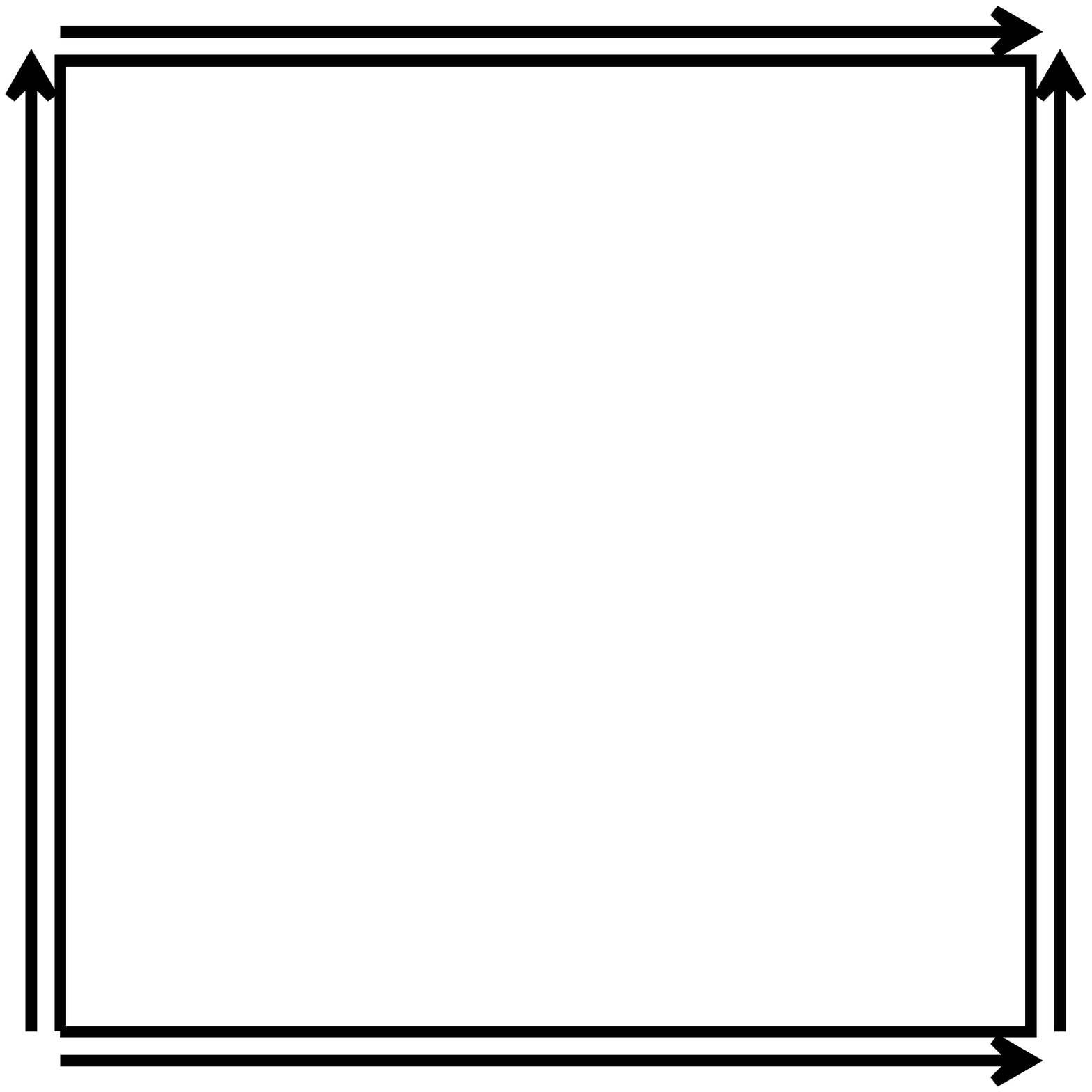}}
\end{minipage}
\hfill
\begin{minipage}[h]{0.49\linewidth}
\center{\includegraphics[width=1.0\linewidth]{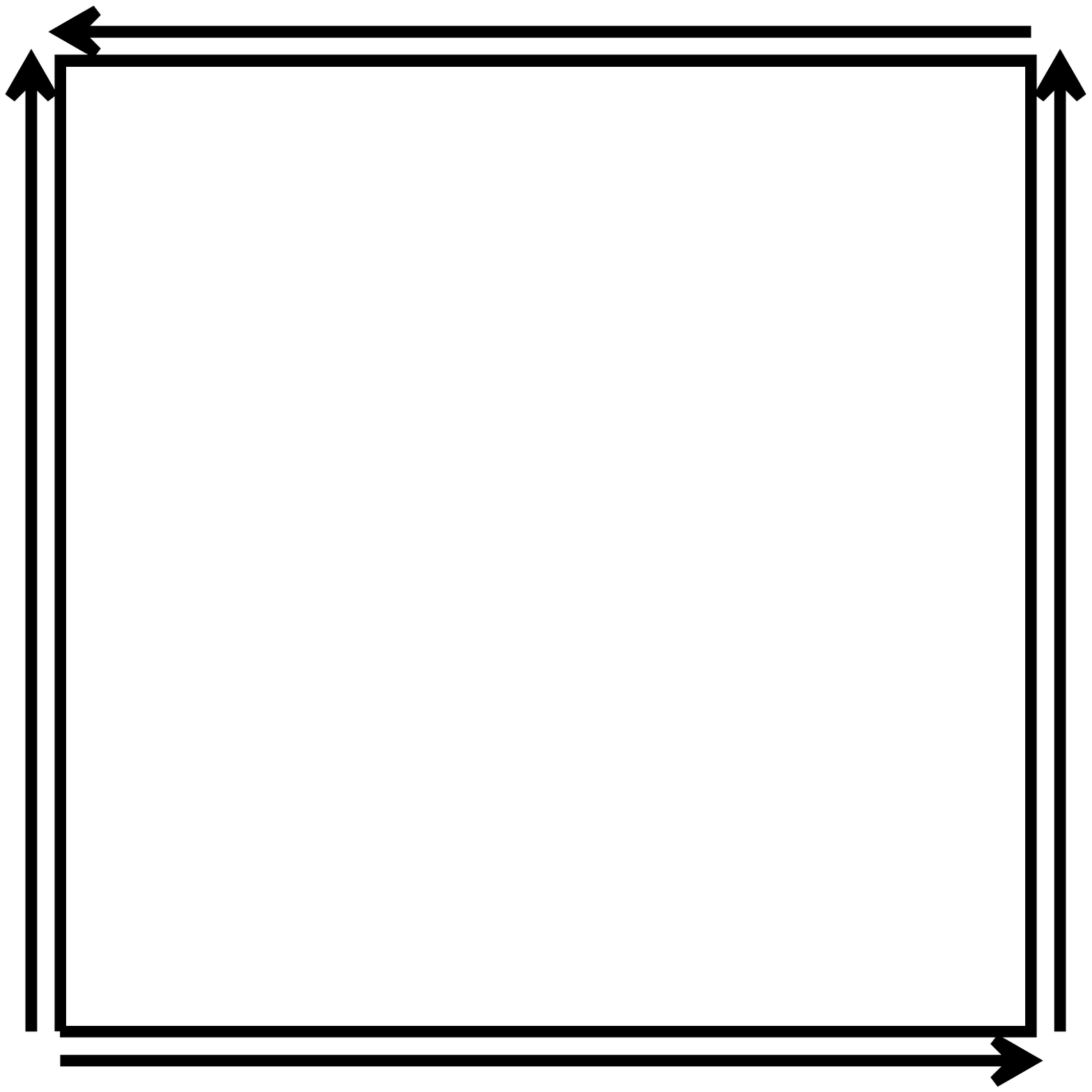}}
\end{minipage}
\caption{Gluing of the boundaries providing the (left) torus and (right) Klein bottle.}
\label{fig:bc}
\end{figure}

Particular attention should be paid to the formulation of boundary conditions, because they determine the topology of the surface on which the calculation is performed. The standard periodic boundary conditions correspond to percolation on the torus: the points of the boundaries in the vertical and horizontal directions are identified as is shown in Fig.~\ref{fig:bc} on the left. This corresponds to the <<gluing>> of a square sheet providing the torus. The gluing according to Fig.~\ref{fig:bc} on the right provides the Klein bottle. In terms of boundary conditions, this means that points on the upper and lower sides of the square are identified in the <<mirror>> manner, which creates additional difficulties for the test of the percolation conditions.

In order to verify the percolation conditions in each of the directions, we used the method proposed in~\cite{machta} adapted for continuum percolation~\cite{perc:2dcont}; the essence of this method is as follows. Each addition of a new disk to the system is accompanied by the procedure of joining of clusters. If the added disk (disk 1) overlaps with another disk (disk 2) and both disks belong to one cluster, there are two paths to their common root disk: through disk 1 and through disk 2. If the sum of these paths in each of the directions is no more than unity, percolation does not occur. If the sum of these paths in a certain direction is larger than or equal to the size of the system in this direction, the wrapping cluster in this direction appears.

\begin{figure}[h!]
\center{\includegraphics[width=1.0\linewidth]{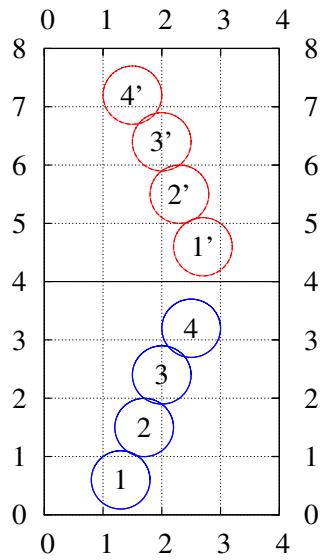}}
\caption{Example of the square calculation domain for $L=4$. Digits and blue color mark disks in the lower square, whereas primed digits and red color indicate their reflections in the upper square with allowance for mirror reflection in the vertical axis.}
\label{fig:example}
\end{figure}

However, it is noteworthy that the described method can be used to verify the percolation conditions only in the case of periodic boundary conditions and percolation on the torus. This method is inapplicable to the Klein bottle because gluing in one of the directions is mirror and the path to the root disk at the transition through the corresponding boundary can change by a value different from $0$ and $\pm 1$. This difficulty is overcome as follows. Instead of the calculation in the square calculation domain with the size $L$, we performed the calculation in a rectangular calculation domain with a vertical size of $2L$ and a horizontal size of $L$, i.e., in a rectangle consisting of two squares in the vertical axis. When adding a new disk, two random numbers $x$, $y$ were generated in the interval $0 \div L$, the disk was added to the lower square of the calculation domain, clusters were joined, and percolation conditions were verified. After that, a <<copy>> of this disk was added to the upper square with allowance for mirror reflection in the vertical axis, i.e., with the coordinates $L-x$ and $y+L$; then, clusters were joined, and the percolation conditions in the axes were verified. Figure~\ref{fig:example} exemplifies the used square calculation domain for the size $L=4$. Digits and blue color mark disks in the lower square, whereas primed digits and red color indicate their reflections in the upper square with allowance for mirror reflection in the vertical axis.

The use of the rectangular calculation domain increases the calculation time with the algorithm as compared to the calculation in the square calculation domain, but eliminates the problem of verifying the percolation conditions, because the rectangular calculation domain has periodic boundary conditions in both axes and makes it possible to applied a method based on the summation of paths to the root disk.

If each calculation ends at the number of disks $n$ that results in the first appearance of the wrapping cluster, the probability of the existence of the wrapping cluster in the canonical ensemble $P_L \left( n \right)$ is equal to the fraction of the calculations ending at the number of disks no more than $n$. The wrapping probability $R_L$ in the grand canonical ensemble is a function of the dimensionless concentration (degree of filling) $\eta = n S/L^2$ ($S$ is the area of one disk) and is obtained by the convolution of the probability $P_L$ with the Poisson distribution with the mean value $\lambda = \eta L^2/S$~\cite{perc:2dcont}:
\begin{equation}
 \label{eq:poisson}
 R_L \left( \eta \right) = e^{-\lambda} \sum_{n=0}^{\infty} \frac{\lambda^n}{n!} P_L \left( n \right) {.}
\end{equation}

At a fixed size of the system $L$, each calculation was performed until percolation occurred in both axes and the number of disks resulting in the first appearance of the wrapping cluster in each of the axes was fixed. Using these data and relation~\eqref{eq:poisson}, we calculated six different probabilities $R_L$ of the appearance of the wrapping cluster:
\begin{enumerate}
 \item In any of the axes $R_L^e$.
 \item In both axes simultaneously $R_L^b$.
 \item In the horizontal axis $R_L^h$ independently of the presence of a cluster in the vertical axis.
 \item In the vertical axis $R_L^v$ independently of the presence of a cluster in horizontal axis.
 \item In the vertical axis, but not in the horizontal axis $R_L^{1v}$.
 \item In the horizontal axis, but not in the vertical axis $R_L^{1h}$.
\end{enumerate}
It is worth noting that the vertical and horizontal axes are not equivalent in contrast to percolation on the torus. Consequently, in the general case, the probabilities are not equal to each other: $R_L^v \ne R_L^h$ and $R_L^{v1} \ne R_L^{h1}$. At the same time, only three of six probabilities are independent because of the obvious relations
\begin{equation}
 \label{eq:prob1}
 R_L^e = R_L^h + R_L^v - R_L^b {,}
\end{equation}
\begin{equation}
 \label{eq:prob2}
 R_L^{1v} = R_L^v - R_L^b {,}
\end{equation}
\begin{equation}
 \label{eq:prob3}
 R_L^{1h} = R_L^h - R_L^b {.}
\end{equation}

The main problem in the simulation of percolation processes is the calculation of the percolation threshold $\eta_c$. The simulation of percolation on the Klein bottle is complicated because the asymptotic values of the probabilities $R_{\infty}$ are unknown in contrast to percolation on the torus. As a result, the relation $R_L \left( \eta_L \right) = R_{\infty}$ cannot be used to determine the dependence $\eta_L \left( L \right)$, which approaches $\eta_c$ in the limit $L \to \infty$.

A method for the determination of the percolation threshold $\eta_c$ based on a nonmonotonic behavior of the functions $R_L^{1v}$ and $R_L^{1h}$ was proposed in~\cite{newmanziff}. The positions of the maxima of the functions $R_L^{1v}$ and $R_L^{1h}$ should approach the percolation threshold $\eta_c$ with an increase in the size of the system $L \to \infty$. Some other authors~\cite{perc:reynolds78,perc:reynolds80} use the expression
\begin{equation}
 \label{eq:etac_12}
 R_L \left( \eta_L \right) = R_{L/2} \left( \eta_L \right) 
\end{equation}
to calculate the critical concentration at a given $L$ value. The sequence of critical concentrations $\eta_L$ thus obtained should converge to the percolation threshold in the limit $L \to \infty$. In this work, we used both methods to estimate the critical concentration.

To test the proposed simulation method, we calculated continuum percolation on the torus with the use of the reflection of the disk placed in the lower square to the upper square. In this calculation, in order to obtain the torus after gluing, reflection in the vertical axis of the disk in the upper square was not performed. The resulting critical concentration $\eta_c$ and probabilities of the percolation transition $R_{\infty}$ are very close to the results of work~\cite{perc:2dcont}.

\section{Results and discussion}

\begin{figure}[h!]
\center{\includegraphics[width=1.0\linewidth]{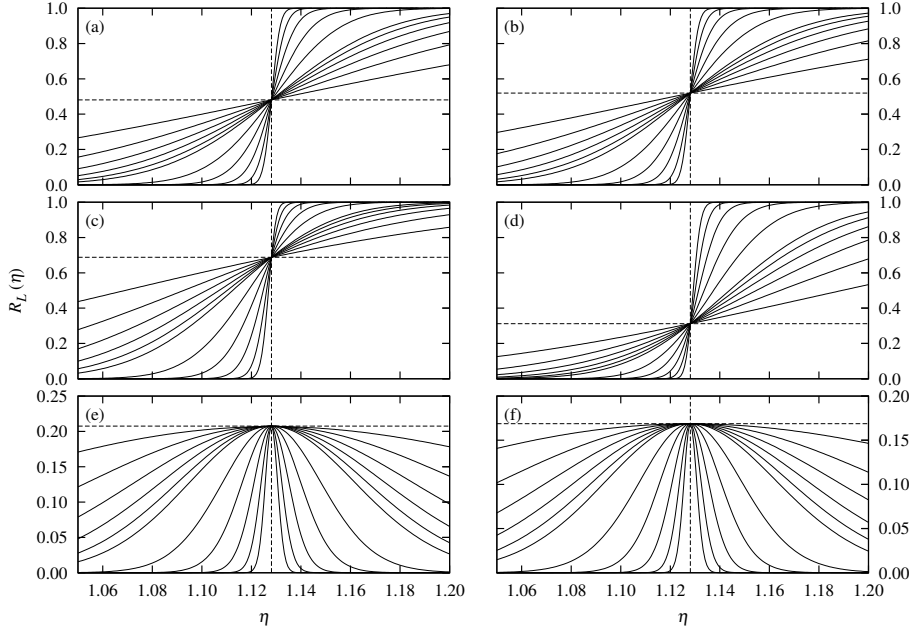}}
\caption{Wrapping probabilities $R_L$ for various sizes of systems from $L=8$ to $L=3072$: (a) $R_L^h$, (b) $R_L^v$, (c) $R_L^e$, (d) $R_e^b$, (e) $R_L^{v1}$, and (f) $R_L^{h1}$. The dashed lines denote the expected values of $\eta_c$ and $R_{\infty}$. }
\label{fig:RL}
\end{figure}

The calculations were performed for 21 sizes of systems in the range from $L=8$ to $L=3072$. For each size, we performed $N$ computer experiments: $N \ge 10^9$ for sizes $8 \le L < 150$, $N \ge 10^8$ for sizes $150 \le L < 600$, $N \ge 10^7$ for sizes $600 \le L < 1200$, and $N \ge 10^6$ for sizes $L \le 1200$. Wrapping probabilities $R_L \left( \eta \right)$ for some sizes from $L=8$ to $L=3072$ are plotted in Fig.~\ref{fig:RL}. It is seen that $R^h_{\infty} \ne R^v_{\infty}$ (Fig.~\ref{fig:RL}a,b) and $R^{v1}_{\infty} \ne R^{h1}_{\infty}$ (Fig.~\ref{fig:RL}e,f) because of the aforementioned nonequivalence of the vertical and horizontal axes in the case of the Klein bottle.

Since the percolation threshold $\eta_c$ for percolation on the Klein bottle is unknown, this threshold was calculated with the use of the dependences $R_L$. In order to determine $\eta_c$, we obtained the dependence of the critical concentration on the size of the system $\eta_L \left( L \right)$, which approaches the percolation threshold in the limit $L \to \infty$. The quantity $\eta_L \left( L \right)$ was determined by two methods: from the position of the maximum of $R_L^{v1}$ and $R_L^{h1}$ and from relation~\eqref{eq:etac_12}.

\begin{figure}[h!]
\center{\includegraphics[width=1.0\linewidth]{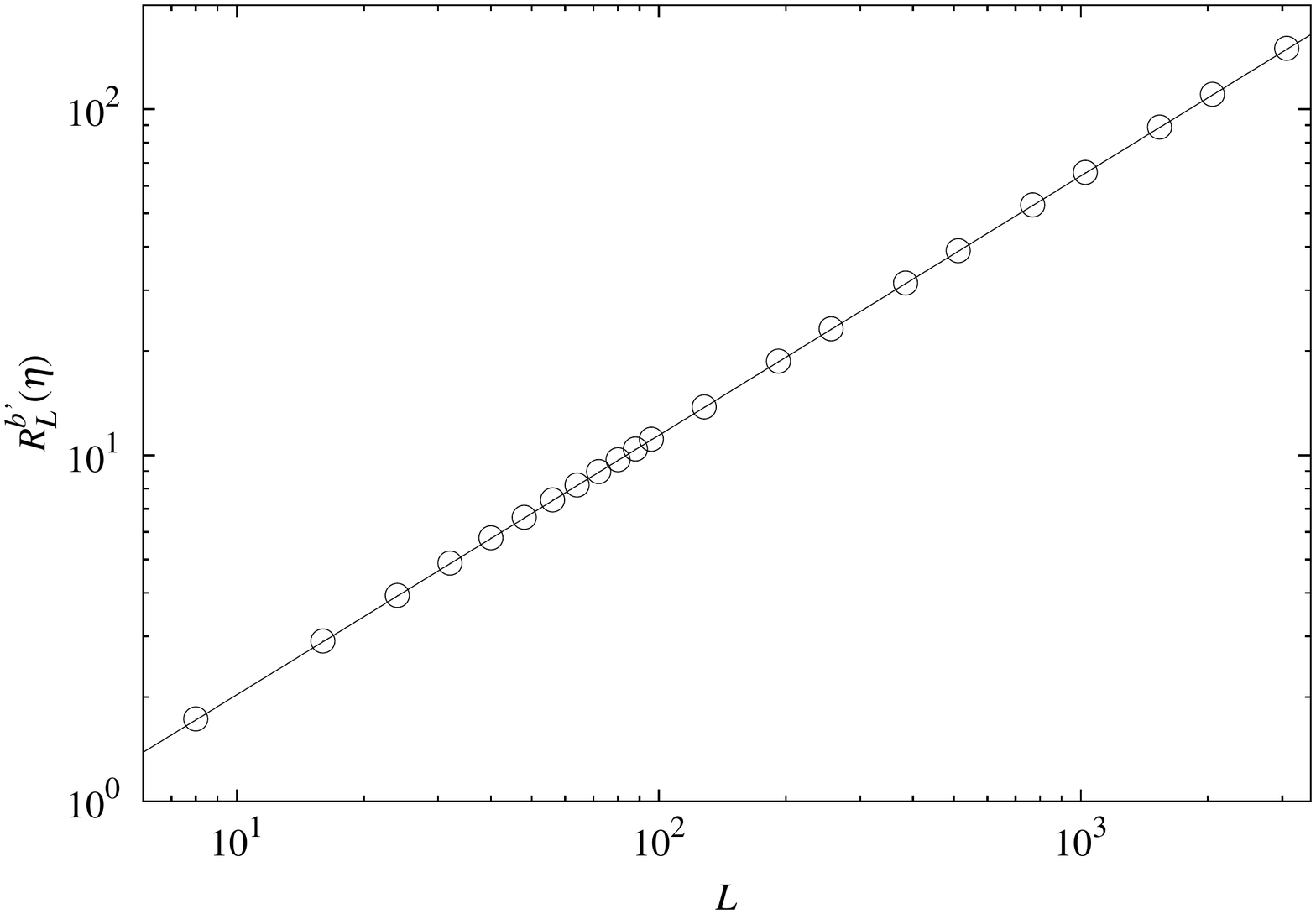}}
\caption{Derivative $R_L^{b'} \left( \eta_L \right)$ versus the size of the system $L$.}
\label{fig:fdiff}
\end{figure}

The rate of convergence $\eta_L$ to the percolation threshold $\eta_c$ is determined by two factors: the characteristic width of the percolation transition region and the rate of convergence of the transition probability $R_L$ to its asymptotic value $R_{\infty}$. The characteristic size of the transition region depends on the size of the system as $L^{-1/\nu}$, where $\nu=4/3$ is the universal critical exponent for two-dimensional percolation. Our calculations show that, in the case of percolation on the Klein bottle, the characteristic size of the percolation transition region, which is determined by the derivative $R_L^{'} \left( \eta_L \right)$, decreases really with an increase in $L$ as $L^{-3/4}$. This is seen in Fig.~\ref{fig:fdiff}, where $R_L^{b'} \left( \eta_L \right)$ is plotted together with the straight line $0.361 L^{3/4}$. Three other derivatives $R_L^{e'} \left( \eta_L \right)$, $R_L^{h'} \left( \eta_L \right)$, and $R_L^{v'} \left( \eta_L \right)$ have the same dependence. 

The rate of convergence of the probability of the percolation transition $R_L$ to $R_\infty$ is determined by the boundary conditions~\cite{newmanziff} and for percolation on the torus depends on $L$ as $L^{-2}$. It will be shown below that the dependence $L^{-2}$~\cite{perc:2dcont} is also valid for percolation on the Klein bottle. Thus, the rate of convergence of the critical concentration $\eta_L$ to the percolation threshold $\eta_c$ in the case of percolation on the Klein bottle is determined by the dependence $L^{-11/4}$.

\begin{figure}[h!]
\center{\includegraphics[width=1.0\linewidth]{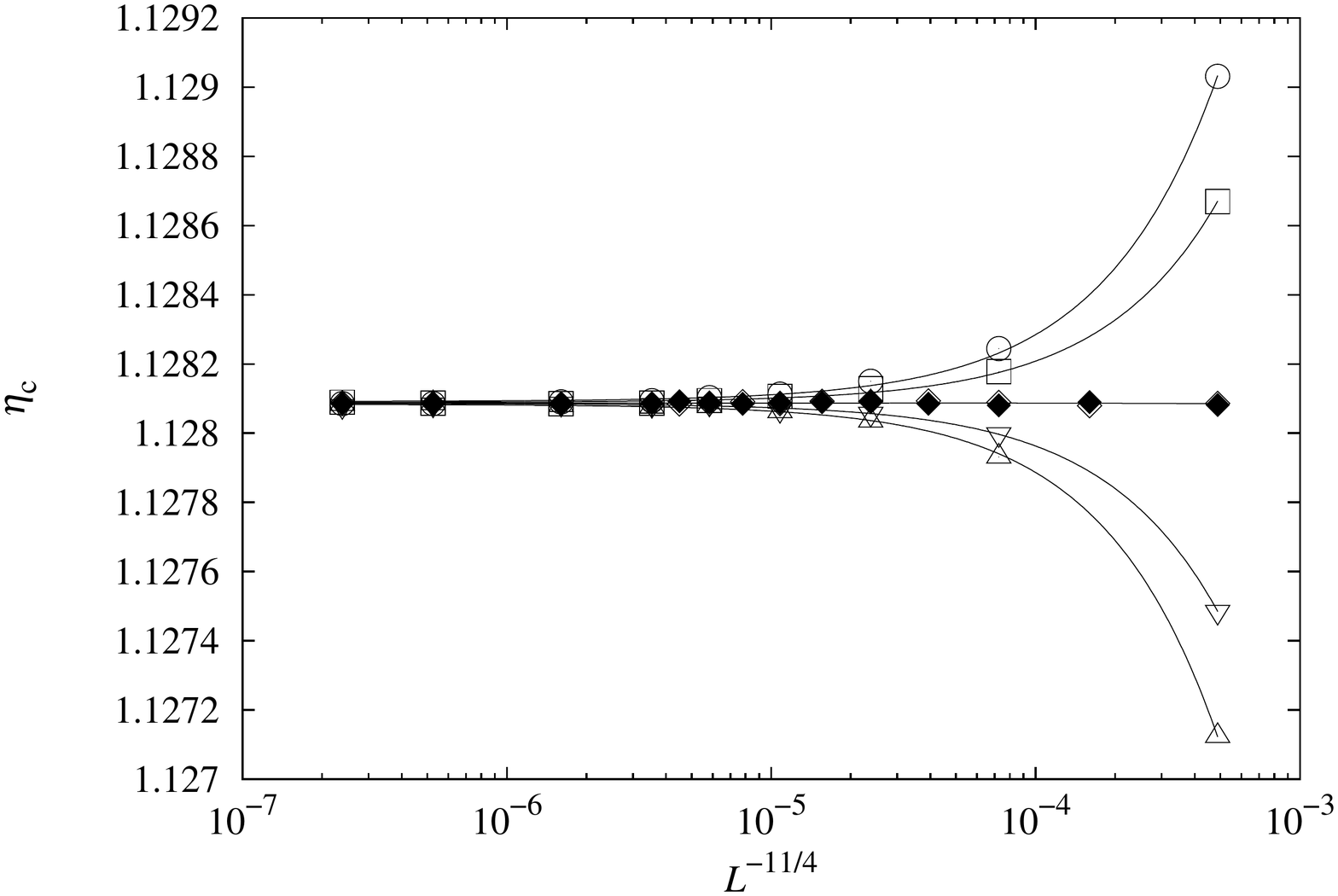}}
\caption{Estimated critical filling factors for continuum percolation on a Klein bottle, derived from the maxima of $R_L^{h1}$ (open diamonds), $R_L^{v1}$ (closed diamonds) and from~\eqref{eq:etac_12} for $R_L^e$ (upward pointing triangles), $R_L^b$ (circles), $R_L^h$ (squares), and $R_L^v$ (downward pointing triangles). Lines are $\eta_L \sim L^{-11/4}$.}
\label{fig:pc}
\end{figure}

The semi-log plots of the critical concentrations $\eta_L \left( L \right)$ derived from the maxima of $R_L^{h1}$ (open diamonds), $R_L^{v1}$ (closed diamonds) and from~\eqref{eq:etac_12} for four probabilities: $R_L^e$ (upward pointing triangles), $R_L^b$ (circles), $R_L^h$ (squares), and $R_L^v$ (downward pointing triangles) are shown in Fig.~\ref{fig:pc}. It is seen that all dependences with an increase in the size approach the same value. The percolation threshold obtained in this work is $\eta_c = 1.128087(1)$, which almost coincides with the percolation threshold in the case of continuum percolation of permeable disks on the torus~\cite{perc:2dcont}. This indicates that the percolation threshold $\eta_c$ is independent of the topology of the surface and is a topological invariant.

The situation with the asymptotic probabilities of the percolation transition $R_{\infty}$ is different. These probabilities for percolation on the torus were calculated in~\cite{perc:pinson} from the conformal field theory and were confirmed in the numerical calculations~\cite{newmanziff,perc:2dcont} for continuum and lattice percolation:
\begin{equation*}
 R^h_{\infty} = R^v_{\infty} = 0.521058289 ... {,}
\end{equation*}
\begin{equation*}
 R^e_{\infty} = 0.690473724 ... {,}
\end{equation*}
\begin{equation*}
 R^b_{\infty} = 0.351642853 ... {,}
\end{equation*}
\begin{equation*}
 R^{h1}_{\infty} = R^{v1}_{\infty} = 0.169415435 ... {.}
\end{equation*}
These relations are invalid for percolation on the Klein bottle. To calculate the $R_{\infty}$ values, we used the following procedure. Taking into account the known percolation threshold $\eta_c$, we calculated the values $R_L \left( \eta_c \right)$ and estimated the asymptotic value $R_{\infty}$ from this dependence by the least squares method with the function $\frac{a}{L^c}+R_{\infty}$ with the unknown parameters $a$, $c$, and $R_{\infty}$. Among these parameters, the scaling coefficient $c$ presenting the rate of convergence is of interest in addition to $R_{\infty}$. It is noteworthy that the functions $R_L$ in the case of percolation on the torus satisfy a power law with an exponent of --2; i.e., $c=2$. The results of the calculations are summarized in the table~\ref{tab:res}.

\begin{table}[h!]
 \caption{Results of the calculations of the asymptotic values $R_{\infty}$ by the least squares method with the function $\frac{a}{L^c}+R_{\infty}$. The rightmost column presents the $R_{\infty}$ values for continuum percolation on the torus~\cite{newmanziff, perc:pinson}.}
 \label{tab:res}
 \begin{center}
 \begin{tabular}{|c|c|c|c|}
 \hline
 type & $R_{\infty}$ & c & $R_{\infty}^{torus}$ \\
 \hline
 $R^h_{\infty}$ & 0.48064(7) & 1.9(0) & \multirow{2}{*}{0.521058...} \\
 \cline{1-3}
 $R^v_{\infty}$ & 0.51934(9) & 1.9(8) & \\
 \hline
 $R^e_{\infty}$ & 0.68813(8) & 1.8(8) & 0.690473... \\
 \hline
 $R^b_{\infty}$ & 0.31185(8) & 1.8(3) & 0.351642... \\
 \hline
 $R^{h1}_{\infty}$ & 0.16878(9) & 1.7(2) & \multirow{2}{*}{0.169415...} \\
 \cline{1-3}
 $R^{v1}_{\infty}$ & 0.20749(1) & 1.8(9) & \\
 \hline
\end{tabular}
 \end{center}
\end{table}

\begin{figure}[h!]
\center{\includegraphics[width=1.0\linewidth]{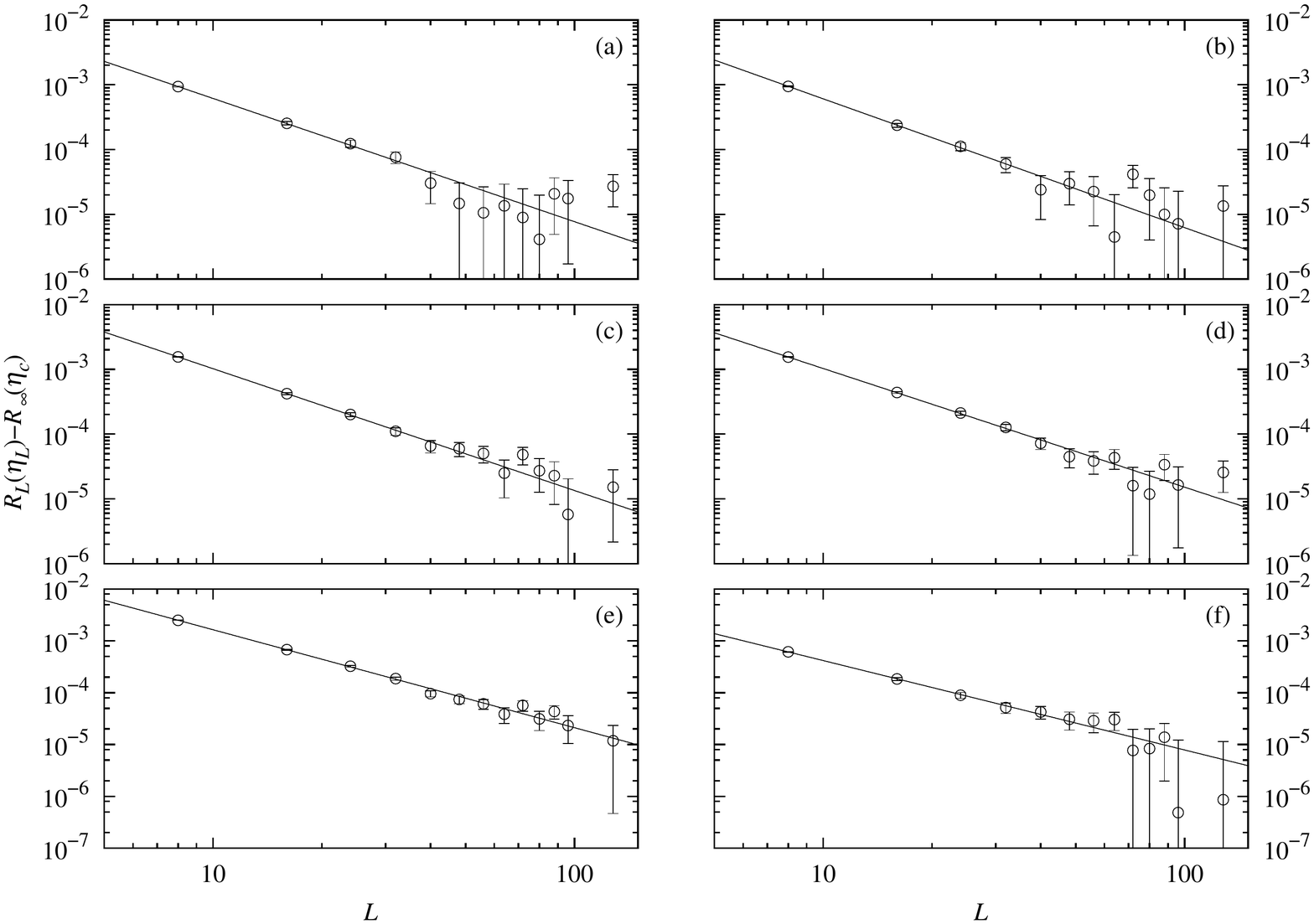}}
\caption{Difference $R_L - R_{\infty}$ versus the size of the system $L$ for $R^h$ (a), $R^v$ (b), $R^e$ (c), $R^b$ (d), $R^{v1}$ (e), and $R^{h1}$ (f). The solid lines are the dependences $L^{-c}$ with the exponents $c$ from the table~\ref{tab:res}.}
\label{fig:RLd}
\end{figure}

It is seen that the exponents $c$ for all types of the probabilities are close to 2. This means that the scaling dependence of the probabilities $R_L$ is independent of the topology, which confirms the above conclusion that the critical concentration $\eta_L$ tends to the percolation threshold $\eta_c$ according to the law $L^{-11/4}$. This law is independent of the topology of the surface. The rate of convergence of the dependences $R_L$, which is determined by the coefficient $c$, is illustrated in Fig.~\ref{fig:RLd}, which shows the dependences of the absolute value of the difference $R_L - R_{\infty}$ on the size of the system. The errors on the plots were calculated from the relation~\cite{perc:2dcont}
\begin{equation}
 \label{eq:sigma}
 \sigma_{R_L} = \sqrt{\frac{R_L \left( \eta \right) \left( 1 - R_L \left( \eta \right) \right) }{N}} {.}
\end{equation}

It is worth noting that the asymptotic values of the probabilities for percolation on the Klein bottle are smaller than those for percolation on the torus~\ref{tab:res}. Moreover, the probabilities $R^h_{\infty}$ and $R_{\infty}^{h1}$ do not coincide with the probabilities $R^v_{\infty}$ and $R_{\infty}^{v1}$ because the vertical and horizontal axes are not equivalent in the case of the Klein bottle. We also note that the wrapping probability for the vertical axis on the Klein bottle $R^v_{\infty}$ is closer to the wrapping probability on the torus than the wrapping probability for the horizontal axis $R^h_{\infty}$. It should be reminded that the standard <<gluing>> without reflection corresponding to periodic boundary conditions is used for the horizontal axis, whereas <<gluing>> with reflection is applied for the vertical axis. At the same time, the probability $R^{h1}_{\infty}$ is closer to the value on the torus than the probability $R^{v1}_{\infty}$.

\section{Conclusions}

The percolation threshold $\eta_c$ and asymptotic values of the wrapping probabilities $R_{\infty}$ for the continuum percolation of permeable disks on the Klein bottle, as well as scaling exponents determining the contribution of finite-size effects to the critical concentration $\eta_L$ and wrapping probabilities $R_L$, have been obtained. It has been shown that the percolation threshold $\eta_c$ and scaling exponents are independent of the topology of the surface and coincide with the respective values obtained for continuum percolation on the torus. At the same time, the wrapping probabilities differ from the respective values on the torus. In addition, since the vertical and horizontal axes are not equivalent in the case of percolation on the Klein bottle, $R^h_{\infty} \ne R^v_{\infty}$ and $R^{h1}_{\infty} \ne R^{v1}_{\infty}$. In this case, the probability $R^v_{\infty}$ is closer to the value on the torus than the probability $R^h_{\infty}$, whereas the probability $R^{v1}_{\infty}$ is farther from the probability on the torus than the probability $R^{h1}_{\infty}$. The difference between the probabilities $R_{\infty}$ for the torus and Klein bottle means that the percolation cluster on one of these surfaces (or on both surfaces) is degenerate.

\section{Acknowledgments}

This work was supported by the state targets higher education institutions in 2015 and the planning period of 2016 in terms of R\&D [project code: 3.720.2014/K] and by the Russian Foundation for Basic Research (project nos. 14-08-00895a and 14-08-00805a). The calculations were partially performed at the <<Basov>> Unit, Computer Cluster, National Research Nuclear University MEPhI.

\end{document}